\documentclass[a4paper]{mem}
\usepackage{natbib}
\usepackage{graphicx}
\usepackage[a4paper]{hyperref}
\idline{1}{1}
\begin{document}
   \title{Ground--based characterisation of the asteroseismic
targets for the COROT space mission
}

   \author{E. Poretti}

   \offprints{E. Poretti}
\mail{via E. Bianchi 46, 23807 Merate (LC)}

   \institute{INAF--Osservatorio Astronomico di Brera,
via E.~Bianchi 46, 23807 Marate (LC) \email{poretti@merate.mi.astro.it}\\ }

   \abstract{We illustrate the approach to the use of asteroseismology
to sound stellar interiors. We also present  
   the spectroscopic and photometric observations carried out to give the
complete characterisation of the potential targets of the space mission COROT. 
The Italian contribution plays an important r\^ole in the preparation
of the observational basis of this space mission.
   }
   \authorrunning{E. Poretti}
   \titlerunning{Ground--based observations for COROT}
   \maketitle
%

\section{Introduction}
The COROT ({\it COnvection, ROtation and planetary
Transits})
mission has been scheduled by the French
space agency, CNES, to be launched in late 2005. The final decision 
confirming the mission has been recently taken (April 2003). Besides a major French
participation (80\%), the COROT mission also involves other european countries:
Spain, Austria, Belgium, Germany as well as a participation of ESA. In  a
short--sighted way,
the Italian Space Agency ASI withdrew the official participation to the project.
 Despite this, a group of researchers still remains active on
various aspects of the project.

The COROT mission is intended to provide high-precision (micro-magnitude level) photometric
monitoring of stellar targets to achieve three main objectives:
\begin{enumerate}
\item  stellar seismology of
dwarf stars to give direct information on the structure and dynamics of
their interiors; among those, a dozen bright stars (m$_V \leq 6.5$) will
be studied continuously for 150 days (central
program, noise level in the frequency spectrum 0.7 ppm over a 5 d time baseline),
 while about 100 fainter (m$_V \leq 8$) stars will be monitored for
up to 30 days in an exploratory phase of the mission (exploratory program;
noise level in frequency spectrum 2.5 ppm over a 5 d time baseline)
once they are selected on the basis of the ground-based preparatory survey;
\item the detection of Earth-like planets from eclipses of their parent stars;
\item the  accurate high precision, continuous,
photometric monitoring of many thousands of fainter stars that lie in the
target fields.
\end{enumerate}
The orbital plane of the satellite will be not subjected to precession
so that all observations should be made in the same area, defined by a cone
with 11$^{\circ}$ half-angle, except for a
180$^{\circ}$ rotation.
More precisely, the targets of the seismology central program are bright F--G
dwarfs and $\delta$~Scuti stars, while those of the exploratory program are
fainter stars, and will be chosen so as to cover the HR
diagram near the Main Sequence, as completely as possible.
\section {The asteroseismic aspect}
The COROT mission should detect a large number of excited modes
for each target.
In stars hotter than the Sun,  the oscillations are driven by opacity. They are found in the
instability strip and amongst more massive and hotter stars on the main
sequence. 
Each frequency is associate with a nonradial mode pattern ($\ell, m)$
on the surface.
We remind the reader that $\ell$ determines the horizontal wavelength 
(the number of patches on the
surface), while $m$ measures the number of nodes around the equator.
For each spherical harmonic, the star supports a series of modes of different
structure in the radial direction, characterized by the radial order $n$,
which, approximately, measures the number of radial nodes. For a spherically
symmetric star, the oscillation frequencies do not depend on $m$. This
degeneracy is lifted by any departure from spherical symmetry; in particular,
rotation induces a splitting with respect to $m$.

Observations of stellar oscillations provide information on many aspects of the
stellar interior. The frequencies of the oscillations depend on the structure
of the star, particularly the distribution of sound speed and density, and on
gas motion and other properties of the stellar interior. The amplitudes of the
oscillations are determined by the excitation and damping processes, which may
involve turbulence from convection, opacity variation and magnetic fields. The
best targets are stars which oscillate in several modes simultaneously. Each
mode has a slightly different frequency, reflecting spatial variations of the
structure within the star, and the combination places strong constraints on
the internal properties.

Although this means we cannot extract the same level of detail on sound speed as
we did for the Sun, 
observations of stellar oscillations provide information on many aspects of the
eigenspectrum according to the asymptotic pulsation theory.  In such
an eigenspectrum, the frequency $\nu$  of  a mode of degree $\ell$ and overtone $n$ is
\begin{equation}
\nu_{nl}\sim\Delta\nu_0(n + \ell/2) + \delta(\ell(\ell+1)\Delta\nu_0^2)/\nu + ...
\end{equation}
The second term in this equation is small: at a first approximation, the $p$-mode 
eigenspectrum will be a combination of frequencies with nearly equal spacing $\Delta\nu_0$
since the 
modes with  $(\ell, n)$ and $(\ell\pm2, n\pm1)$ are almost degenerate in frequency. The
fundamental frequency spacing $\Delta\nu_0$ is equal to the time for a sound wave to
traverse the diameter of the star. This depends on the square root of the star's mean
density, so is sensitive to the stellar mass and radius.
Actually, the second term in the equation is not zero, so the mode degeneracy is broken.
The magnitude of the second--order splittings, $\delta$, depends heavily on the sound
speed gradient in the star's core. In the nearly isothermal core of a star, that
gradient depends most strongly on the composition gradient created by fusion reactions.
This gradient changes with stellar age, as the star gradually converts its central
supply of hydrogen into helium. Therefore, the second-order splittings of the
eigenspectrum act as a main--sequence lifetime clock.

Rotation may have important effects on stellar structure and evolution and it
affects the oscillations, leading to a splitting of the frequencies according
to $m$:
\begin{equation}
\nu_{nlm} = \nu_{nl0} + m \Omega_{nl}/2\pi
\end{equation}
where $\Omega_{nl}$ is an average of the angular velocity weighted by the
structure of the given mode. When only low--degree modes are observed these
averages all extend over  most of the star. However, even such limited
information will give some indication of the variation of rotation with depth
in the star, particularly if it is combined with measurement of the surface
rotation (via spectroscopic observations, for example).

\subsection{The spectroscopic observations}

\begin{table}
\caption{Summary of the spectroscopic observations.}
\begin{tabular} {l r}
\hline
\multicolumn{2}{c}{COROT potential primary}\\
\multicolumn{2}{c}{and secondary targets} \\
\hline
\noalign{\smallskip}
 Total number   &     957    \\      
 In the Galactic Center direction & 366      \\ 
 In the Galactic AntiCenter direction & 591   \\  
 In the Northern Hemisphere &  548\\ 
 In the Southern Hemisphere &  409\\ 
\noalign{\smallskip}
 Observed at  ESO/FEROS& 423 \\
 Observed at  OHP/ELODIE   &  419 \\ 
 Observed at Brasil/FEROS &  87\\
 Observed at TNG/SARG &  73 \\
 Observed at Euler/Coralie & 17 \\
 Observed at Tautenburg &  11 \\
\noalign{\smallskip}
\hline
\label{sol}
\end{tabular}
\end{table}
\begin{figure}
\resizebox{\hsize}{!}{\includegraphics{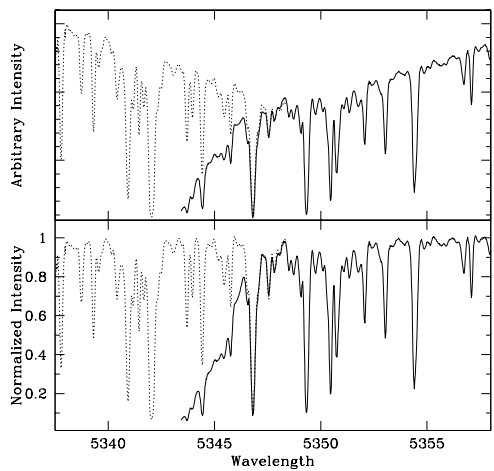}}
\caption{Example of the automatic normalization procedure applied to
the SARG spectrum of HD~44019. Top panel: original data for two adjacent orders,
indicated with different line types.
Bottom panel: normalized data for the same orders. The procedure  
works well, except for the extreme blue border of the order more to the
red, where the intensity drops too fast.} 
\end{figure}

It is essential to have an accurate and reliable knowledge of
T$_{\rm eff}$, log g, abundances, v\,sini parameters for the 
potential targets in order to quickly exploit the photometric
data which will be supplied by COROT.  One of the goals of
the mission is to provide seismological analysis of stars well
distributed in this parameter space. Therefore, one have been selected 957 stars located in
the COROT field--of--view which can be potential primary or secondary targets,
i.e., we excluded stars fainter than $V$=8.0 and known giant stars.
The observational effort to secure at least one high--resolution ($R\sim 50\,000$)
spectrum involved several telescopes and intruments (Tab.~1); the task has been 
achieved in due time (February 2003) and SARG provided an important contribution, allowing to
complete the survey of northern targets. Moreover, ESO/FEROS observations
have an Italian P.I. and the FEROS automatic pipeline has been improved
by the Merate team. A further, original Italian
contribution to COROT is based on the monitoring of some targets 
to establish the level of the stellar activity (Catania team, observations
from Serra La Nave). A team chaired by Werner~W.~Weiss (Vienna University) is
trying to build--up a semi--automatic procedure
to reduce SARG spectra. 
Figure~1 illustrates the application of the current version of
the automatic normalization procedure to HD~44019 (spectral type K2),
one  of the
COROT candidate targets. It is a zoom of a region with two overlapping orders.
As can be seen, the normalization (bottom panel) is satisfactory
compared with the original data (top panel) despite the presence of
a lot of spectral lines.  It is, however,  still unsatisfactory at the extreme
blue border of the order more to the red, where the intensity drops too fast.
Improvements are expected in the near future.

To determine automatically the effective temperature, gravity, and metallicity of
the observed stars from the high--resolution
spectra, the current approach uses primarily the {\sc etoile} programme
(Katz 2000), which
is an extension of the {\sc tgmet} programme developed by Katz et al.
(1999).  A new, semi--authomatic
method has been developed to analyze the high--resolution spectra ({\sc vwa}; 
Bruntt et al. 2002). The procedure
selects the least blended lines from the
atomic database {\sc vald} and consequently adjusts the abundance in
order to find the best match between the calculated and observed
spectra. 
The whole atlas of the high-resolution spectra and all the derived fundamental
parameters  will be made accessible to the community, constituting by their own
a powerful tool to investigate properties of stars along the main sequence.
Teams working on atmosphere models are invited to join this aspect of the
project.

\subsection{The photometric observations}

The $\delta$ Sct class
covers both the early main-sequence evolutionary stage, when the star is
burning hydrogen in the core, and the following one, when it
leaves the Terminal--Age Main Sequence 
burning hydrogen in a shell. These two different
stages correspond to different types of structures and their study offers
different insights into the physics of the stellar interiors.
Therefore, both types of variables deserve interest.
However, the primary objectives of COROT are highly focused on
core overshooting processes and on transport of angular momentum and
chemical species. These processes are crucial in the main--sequence stage
for intermediate-- and high--mass stars. None of the well--known $\delta$
Sct stars studied in the past years is included in the COROT field--of--view.
Therefore, we put a strong priority in searching for new $\delta$ Sct stars
close to the Main Sequence. The observational effort has been primarily addressed
in the Center direction, as there are no good primary targets in such a 
direction. We photometrically surveyed 68 potential $\delta$ Sct
variables  and we found that 23\% of
the sample actually display multiperiodic light variability up to few mmag of
amplitude (Poretti et al. 2003).
We note that by observing a zone of 450~deg$^2$ on or just above the galactic plane in the solar
neighbourhood we discovered variables well
mapping the lower part of the instability strip. Accurate $v\sin i$ values
($\pm$5~km\,s$^{-1}$) have also been obtained on the basis of the spectroscopic
observations collected for the purposes described in the previous subsection.
\section{Conclusions}
The Italian contribution to the COROT space mission will allow 
the full spectroscopic characterisation of the targets
(by means of the high--resolution spectra taken with SARG, FEROS and 
the stellar activity monitoring from Serra La Nave) and
to a precise photometric  evaluation of the $\delta$ Sct variability
in the COROT field--of--view and, more in general, in the lower
part of the instability strip. If we also consider that observations
with adaptive optic (to detect close companions of primary targets)
have also been undertaken, the Italian contribution continues to be both 
original and useful, despite the ASI withdrawal from the project. 
\begin{acknowledgements}
The author wishes to thank Werner W.~Weiss for the data reduction 
(within project P14984 of the FWF Austria) which inspired Fig.~1.
\end{acknowledgements}
\bibliographystyle{aa}

\end{document}